# Charge transport in lithium peroxide:
# Relevance for rechargeable metal-air batteries

[a] Department of Physics, University of Michigan, Ann Arbor, Michigan 48109, United States
[b] Department of Mechanical Engineering, University of Michigan, Ann Arbor, Michigan 48109, United States
* Corresponding author. E-mail: djsiege@umich.edu

(Dated: 13 May 2013)

The mechanisms and efficiency of charge transport in lithium peroxide ($Li_2O_2$) are key factors in understanding the performance of non-aqueous Li-air batteries. Towards revealing these mechanisms, here we use first-principles calculations to predict the concentrations and mobilities of charge carriers and intrinsic defects in $Li_2O_2$ as a function of cell voltage. Our calculations reveal that changes in the charge state of $O_2$ dimers controls the defect chemistry and conductivity of $Li_2O_2$. Negative lithium vacancies (missing $Li^+$) and small hole polarons are identified as the dominant charge carriers. The electronic conductivity associated with polaron hopping ($5 \times 10^{-20}$ S/cm) is comparable to the ionic conductivity arising from the migration of Li-ions ($4 \times 10^{-19}$ S/cm), suggesting that charge transport in $Li_2O_2$ occurs through a mixture of ionic and polaronic contributions. These data indicate that the bulk regions of crystalline $Li_2O_2$ are insulating, with appreciable charge transport occurring only at moderately high charging potentials that drive partial delithiation. The implications of limited charge transport on discharge and recharge mechanisms are discussed, and a two-stage charging process linking charge transport, discharge product morphology, and overpotentials is described. We conclude that achieving both high discharge capacities and efficient charging will depend upon access to alternative mechanisms that bypass bulk charge transport. More generally, we describe how the presence of a species that can change charge state – e.g., $O_2$ dimers in alkaline metal-based peroxides – may impact rechargeability in metal-air batteries.

## Introduction

Thanks to their high theoretical specific energy density, rechargeable non-aqueous Li-air batteries are attracting increasing attention as a future energy storage technology.[1–4] In the absence of undesirable side reactions (*e.g.* degradation of the electrolyte or carbon support[3,1,4]), a Li-air cell can be described by the reversible reaction $2Li + O_2 \leftrightarrow Li_2O_2$. This chemistry is unlike conventional Li-ion intercalation electrodes because the solid phase discharge product, lithium peroxide ($Li_2O_2$), nucleates and grows on the cathode during discharge, and subsequently decomposes during recharge.

In order to achieve a high energy density the cathode of a Li-air cell should be substantially filled with $Li_2O_2$ at the end of discharge. However, prior studies have suggested that charge transport limitations through an ostensibly insulating $Li_2O_2$ discharge phase may constrain the capacity and rate capability of Li-air cells.[5–9] Therefore, a question of both practical and fundamental importance to the Li-air system is the mechanism of charge transport through the discharge product.[10,11] For example, Viswanathan *et al.* have investigated electron tunneling through thin, dense films of $Li_2O_2$ and found that this mechanism cannot support appreciable currents beyond a thickness of ~5 nm.[5] Nevertheless, the high capacities measured in many experiments,[10,11] in conjunction with the observation of large discharge product particles (diameters up to ~1 micron or larger[10,11]) suggests that other charge transport mechanisms are at play.

Unfortunately, an accepted mechanism for charge transport in Li-air cathodes has yet to emerge.[12,5,13–19] First-principles calculations by Hummelshøj *et al.* predicted that a high concentration of lithium vacancies in $Li_2O_2$ will yield *p*-type conductivity associated with a depletion of electrons from the valence band.[12] Subsequent calculations have also predicted *p*-type conductivity at $Li_2O_2$ surfaces[17,18] and at $Li_2O_2$-carbon interfaces.[16] Other studies have predicted that both holes and electrons will become self-trapped in $Li_2O_2$, forming small hole[14] and small electron[15] polarons. Although hole polarons were at first predicted to have very low hopping barriers,[14] a recent study examining the mobilities of these species in more detail has challenged this notion.[19] Furthermore, the nature and concentrations of charge carriers and intrinsic point defects in $Li_2O_2$ have not been reported. Such information is important because the concentrations of these species, when combined with mobilities, relates to the conductivity of bulk $Li_2O_2$, and thus ties directly to the performance of the battery.

As a step towards elucidating the impact and mechanism of charge transport in Li-air cells, here we employ first-principles calculations to predict the conductivity of crystalline $Li_2O_2$. More specifically, the concentrations of all chemically-relevant intrinsic (point) defects in $Li_2O_2$ are evaluated as a function of cell voltage; subsequent calculations are used to assess the mobilities of the dominant charge carriers. To obtain an accurate description of the electronic structure, hybrid functionals[20,21] and many-body perturbation theory (*GW*) methods[22,23] are employed. Our calculations indicate that charge transport in $Li_2O_2$ is mediated by both the migration of negative lithium vacancies, $V_{Li}^-$, corresponding to missing $Li^+$, and the hopping of hole polarons, $p^+$. For ionic transport, the barrier for $V_{Li}^-$ migration, 0.33-0.39 eV, yields an ionic conductivity of ~$4 \times 10^{-19}$ S/cm. The hopping of hole polarons was found to have in-plane and out-of-plane barriers of 0.42 and 0.71 eV, which are comparable to recent DFT+U calculations,[19] yet are much larger than those suggested by previous HSE06 calculations.[14] We predict an intrinsic electronic conductivity of ~$5 \times 10^{-20}$ S/cm, which would classify $Li_2O_2$ as an insulator. During charging, the partial delithiation of $Li_2O_2$ is expected to increase the conductivity, with each overpotential increment of ~0.1 V increasing the conductivity by an order of magnitude. Such an enhancement may explain why Li-air cathodes that have been loaded with purchased $Li_2O_2$ can be recharged at high overpotentials despite the low conductivity of $Li_2O_2$.[24–27] Our results suggest that recharge may occur via a two-stage process, with thin deposits decomposing at low potentials via electron tunneling, and thick deposits decomposing at moderately high potentials via polaron hopping. Therefore, strategies for enhancing bulk transport – or avoiding altogether it in place of transport via other pathways such as surfaces, grain boundaries, amorphous regions, *etc.* – should be explored. More generally, we discuss how the capability for electronic charge transport in metal-air discharge phases can be tied to the presence of a species that can change valence state, such as the $O_2$ dimers in $Li_2O_2$. The presence or absence of such a species could explain why some non-aqueous metal-air chemistries are rechargeable, while others are not.

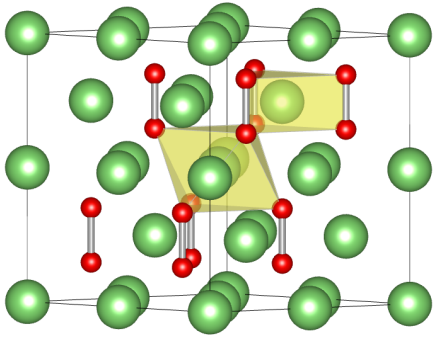

**Figure 1.** Crystal structure of $Li_2O_2$, illustrated using a 2 × 2 × 1 expansion of the unit cell. Large green atoms are lithium, and small red atoms are oxygen. Polyhedra indicate the trigonal prismatic and octahedral coordination of the two unique Li sites.

## Computational methodology

The crystal structure of $Li_2O_2$, shown in Figure 1, consists of alternating layers of trigonal prisms and octahedra/tetrahedra, with oxygen sites lying on the vertices of the polyhedra. One notable feature of the structure is the presence of covalently bonded $O_2$ dimers. As we will describe later, the ability of these dimers to change charge state plays an important role in the defect chemistry and conductivity of $Li_2O_2$. All of the octahedra (O) and half of the trigonal prisms (TP) are occupied by lithium atoms.

Point defect formation energies were calculated for 23 unique species, including vacancies, divacancies, interstitials, polarons, and bipolarons. First principles calculations were performed using the Vienna *ab initio* simulation package (VASP)[28–31] with a 3×3×2 (144-atom) supercell. Makov-Payne finite-size monopole corrections[32,33] and oxygen overbinding corrections[34–36] were included. See Supplementary Information for details. Given that self-interaction errors inherent to semilocal functionals (*e.g.* GGAs) can lead to qualitatively incorrect descriptions of certain defects,[37–39] our calculations employ the HSE hybrid functional[20,21] in conjunction with many-body perturbation theory (GW) methods.[22,23] The incorporation of exact exchange in the HSE family of functionals serves to compensate for self-interaction error, with the mixing parameter $\alpha$ determining what fraction of the semilocal exchange is replaced with exact exchange. As is common practice, we have adjusted $\alpha$ to reproduce the bandgap from GW-based methods.[39,38,40] This approach is motivated by the fact that the positions of the band edges are important for determining defect properties. We find that a mixing parameter of $\alpha = 0.48$ yields a gap of 6.63 eV, which agrees well with the reference gap of 6.73 eV derived from an average of $G_0W_0$ and scGW calculations on bulk $Li_2O_2$ (see Supplementary Information for details). As discussed below, we also investigated the sensitivity of our results to the choice of mixing parameter.

The equilibrium concentration $C$ of a defect $X$ in charge state $q$ in a given solid phase can be written as $C(X^q) = D_X e^{-E_f(X^q)/k_B T}$, where $D_X$ is the number density of defect sites.[41] Defect formation energies $E_f$ are calculated according to:[33]

$$E_f(X^q) = E_0(X^q) - E_0(\text{bulk}) - \sum_i n_i \mu_i + q\varepsilon_F + E_{MP1}$$

where $n_i$ is the number of atoms of the $i^{th}$ species in the defect, $\mu_i$ is the chemical potential of that species, $\varepsilon_F$ is the Fermi level, and $E_{MP1}$ is the Makov-Payne monopole size correction.[32,33] Size convergence tests are reported in the Supplementary Information. The chemical potential of oxygen was assumed to be fixed by equilibrium with oxygen in the atmosphere, while that of lithium was set by ion exchange with the anode:[12]

$$\mu_{Li}(\text{cathode}) = \mu_{Li}(\text{BCC Li}) - eE$$

where $E$ is the cell voltage. Additional details are provided in the Supplementary Information.

## Results

### Defect formation energies

Figure 2 shows the formation energies for the low-energy defects as a function of the Fermi level for isolated $Li_2O_2$ (or equivalently, a cell whose potential is at the open circuit voltage). Table 1 summarizes the equilibrium formation energies and concentrations for all defects examined. As shown in Figure 2, the dominant (i.e., lowest energy) positively charged species is the hole polaron, $p^+$. The hole polaron consists of a hole that is self-trapped at an oxygen dimer, reducing the formal charge on a peroxide ($O_2^{2-}$) dimer by one to yield a superoxide ($O_2^{1-}$) dimer and an associated contraction of the covalent O-O bond.[14,19] The dominant negative defect species is the negative lithium vacancy (*i.e.* absence of a $Li^+$ ion). As shown in Table 1, negative lithium vacancies at the two symmetry-distinct Li sites have similar energies, with $V_{Li}^-$ (TP) being only 20 meV more stable than $V_{Li}^-$ (O). The concentrations of the dominant charge carriers, $p^+$ and $V_{Li}^-$, are established by an overall charge neutrality condition, and have values of $1 \times 10^7$ cm$^{-3}$, which is approximately three orders of magnitude less than the intrinsic carrier concentration in silicon at 300 K (~$10^{10}$ cm$^{-3}$).[42] To quantify the influence of the mixing parameter, we also performed calculations using the "standard" $\alpha$ value of 0.25 (*i.e.* the HSE06 functional); this altered the equilibrium defect formation energies by only ~0.1 eV or less as shown in the Supplementary Information. The influence of the mixing parameter is discussed in more detail below.

**Table 1.** Equilibrium defect formation energies (eV) and concentrations (cm$^{-3}$) in $Li_2O_2$.

| | | | |
|---|---|---|---|
| $p^{2-}$ | 3.12 (1 × 10$^{-30}$) | $V_{O2}^-$ | 2.47 (1 × 10$^{-19}$) |
| $p^-$ | 1.51 (1 × 10$^{-3}$) | $V_{O2}^0$ | 4.71 (2 × 10$^{-57}$) |
| $p^+$ | 0.95 (1 × 10$^7$) | $V_{O2}^+$ | 4.32 (1 × 10$^{-50}$) |
| $V_{Li}^-$ (O) | 0.95 (3 × 10$^6$) | $V_{O2}^{2+}$ | 3.24 (9 × 10$^{-33}$) |
| $V_{Li}^-$ (TP) | 0.93 (7 × 10$^6$) | $O_i^{2-}$ | 4.55 (4 × 10$^{-54}$) |
| $V_{Li}^0$ (O) | 1.37 (4 × 10$^{-1}$) | $O_i^-$ | 4.34 (1 × 10$^{-50}$) |
| $V_{Li}^0$ (TP) | 1.02 (2 × 10$^5$) | $O_i^0$ | 1.33 (5 × 10$^0$) |
| $V_{Li}^+$ (O) | 2.05 (1 × 10$^{-12}$) | $O_i^+$ | 2.22 (5 × 10$^{-15}$) |
| $V_{Li}^+$ (TP) | 1.45 (1 × 10$^{-2}$) | $Li_i^-$ | 3.80 (1 × 10$^{-41}$) |
| $V_O^-$ | 3.58 (4 × 10$^{-38}$) | $Li_i^0$ | 2.51 (6 × 10$^{-20}$) |
| $V_O^0$ | 0.74 (2 × 10$^{10}$) | $Li_i^+$ | 1.69 (1 × 10$^{-6}$) |
| $V_O^+$ | 1.66 (9 × 10$^{-6}$) | | |

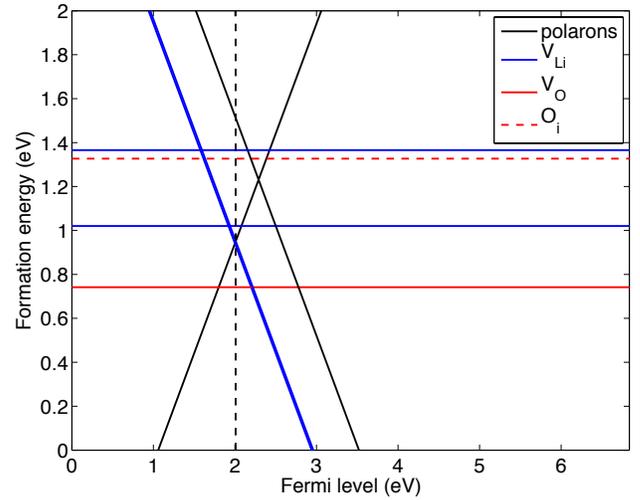

**Figure 2.** Formation energies of low-energy defects in $Li_2O_2$. Positive defects have an upwards slope while negative defects have a downwards slope. The vertical dashed line indicates the position of the Fermi level that satisfies charge neutrality.

Figure 2 also shows that the neutral oxygen vacancy is the most stable uncharged defect, with a formation energy of 0.74 eV. At first glance such a low formation energy may seem surprising because the creation of an oxygen vacancy requires the cleavage of an oxygen-oxygen bond. However, this cleavage results in the reduction of the remaining oxygen ion to a -2 charge state, which is energetically favorable. The second lowest energy neutral defect is the neutral lithium vacancy, $V_{Li}^0$ ($E_f$ = 1.02 & 1.37 eV for the two Li sites), which consists of a $p^+$-$V_{Li}^-$ bound pair. The binding energy $\Delta E = E_f(p^+) + E_f(V_{Li}^-) - E_f(V_{Li}^0)$ is 0.53 and 0.86 eV at the O and TP sites. A previous study[14] suggested that a hole polarons in $Li_2O_2$ would be bound to lithium vacancies on the basis that the $p^+$-$V_{Li}^-$ binding is fairly strong. However, as can be seen from Table 1, the equilibrium concentrations of unbound $p^+$ and $V_{Li}^-$ are in fact higher than that of $V_{Li}^0$ due to the entropy gain associated with dissociation.[41]

### Defect mobilities

Conductivity in $Li_2O_2$ can in principle arise from the migration of charged defects (ionic conductivity) and/or hopping of small polarons (electronic conductivity). In both cases the conductivity depends on the concentrations and mobilities of these species, and can be expressed as:[43]

$$\sigma = \frac{Cva^2e^2}{k_BT} e^{E_b/k_BT},$$

where $C$ is the concentration, $v$ is a hopping attempt rate that we take to be $10^{13}$ s$^{-1}$,[44,15] $a$ is the hop length, and $E_b$ is the hopping barrier.

We first consider the ionic conductivity associated with $V_{Li}^-$ migration. Energy barriers for five migration pathways were calculated using the nudged elastic band (NEB) method (see Supplementary Information).[45] Because these calculations are computationally expensive, we optimized the migration pathway using the PBE GGA functional and report the barrier obtained at this level of theory;[46] this choice is justified by the fact that the unrelaxed barriers obtained with PBE were essentially the same as the unrelaxed barriers obtained with HSE, indicating that there is little sensitivity to the choice of functional. The lowest energy pathway corresponds to migration between adjacent octahedral and trigonal prismatic sites, with a barrier of 0.33 eV relative to the octahedral site and 0.39 eV relative to the trigonal prism site. Similar values have been found in prior calculations.[12,13] Setting $E_b$ to the average of these two values yields an ionic conductivity of 9 × 10$^{-19}$ S/cm and a defect diffusion coefficient of $D_i = va^2 e^{E_b/k_BT}$ = 6 × 10$^{-9}$ cm$^2$/s. Because this pathway allows for both in-plane and out-of-plane transport, the ionic conductivity is expected to be more or less isotropic.

Next we consider the electronic conductivity associated with hole polarons. In this case we evaluate the energy barrier associated with nearest neighbor hole polaron hopping. While previous studies treated all in-plane (i.e., within a basal plane) hopping paths as symmetry equivalent and all out-of-plane paths as symmetry equivalent,[14,19] a Jahn-Teller distortion due to the degeneracy of $\pi_x^*$ and $\pi_y^*$ molecular orbitals breaks this symmetry. This distortion lowers the polaron's symmetry from $D_{3h}$ to $C_{2v}$ and lowers the ground state energy by 22 meV. As a result of this symmetry breaking there are six symmetry inequivalent in-plane and four symmetry inequivalent out-of-plane paths, as well as a trivial in-place rotation (see Supplementary Information). We calculated the adiabatic barrier for these paths based on a chain of linearly interpolated images.[14,15] We found that all of the in-plane paths had similar barriers, and all of the out-of-plane paths had similar barriers (see Supplementary Information). Attempts to optimize the geometry with the NEB method did not lead to significant changes in the barrier height: after 189 optimization steps, the barrier height of the lowest energy in-plane path was reduced by only 0.04 eV. Figure 3 compares the energy profiles for the lowest energy in-plane and lowest energy out-of-plane hopping paths, for which we find barriers of 0.42 and 0.71 eV, respectively. These values correspond to conductivities of $5 \times 10^{-20}$ and $1 \times 10^{-24}$ S/cm for in-plane and out-of-plane transport. To place the calculated conductivities in context, we note that the conductivity of other battery materials can be orders of magnitude higher: for example, in LiFePO$_4$ $\sigma \sim 10^{-9}$ S/cm,[47] while the conductivity of a good insulator such as fused silica is similar to our predicted value for Li$_2$O$_2$.[48]

Unlike lithium vacancy migration, hole polaron hopping is predicted to be anisotropic, with in-plane transport being favored over out-of-plane transport. It has previously been suggested that this anisotropy could contribute to anisotropies in the morphology of Li-air discharge products.[19] Combining the lithium vacancy defect diffusion coefficient $D_i$ with the hole polaron defect diffusion coefficient ($D_p = 9 \times 10^{-10}$ and $2 \times 10^{-14}$ cm$^2$/s for in-plane and out-of-plane transport), we obtain a chemical diffusion coefficient[49] of $\tilde{D} = 2D_iD_p/(D_i + D_p) = 2 \times 10^{-9}$ and $4 \times 10^{-14}$ cm$^2$/s for in-plane and out-of-plane diffusion.

Regarding experiments, a recent study measuring the ionic and electronic conductivities of Li$_2$O$_2$ arrived at qualitatively the same picture presented here: electronic conduction is mediated by hole polarons, and ionic conduction is mediated by negative lithium vacancies.[50] However, because the experimental sample was in the extrinsic regime – where defect concentrations are controlled by the presence of impurities – the measured electronic and ionic conductivities (at 100 °C) of $10^{-12}$-$10^{-11}$ S/cm and $10^{-10}$-$10^{-9}$ were significantly larger than those predicted here. Consequently, a direct comparison between experimental values and our calculations is not possible.

Another recent experimental study employed $in\ situ$ TEM and found that the Li-O$_2$ discharge product decomposed at the carbon interface, but not the solid electrolyte interface, indicating that electronic charge transport is slower than Li ion transport.[51] This agrees with our calculations, which predict $V_{Li}^-$ to be 18 times more mobile than hole polarons, based on the faster in-plane hopping rate. Furthermore, this study found that recharge of particles of radius $L \approx 200$ nm began at $t < 200$ s when a voltage of $V < 10$ V was applied across the particles. Applying dimensional analysis, we can estimate a lower bound on the mobility of lithium defects of $L^2/Vt = 2 \times 10^{-13}$ cm$^2$/V·s. This is consistent with our calculations, which predict a $V_{Li}^-$ mobility of $\mu = eD_i/k_BT = 2 \times 10^{-7}$ cm$^2$/V·s.

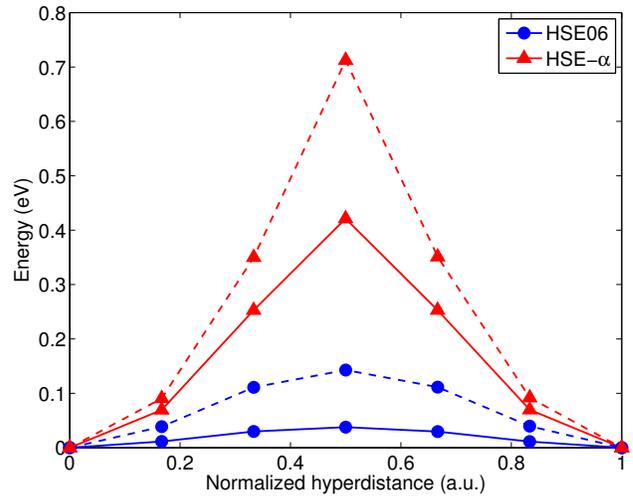

**Figure 3.** Energy profiles for in-plane (solid lines) and out-of-plane (dashed lines) hole polaron hopping.

Finally, a third study has estimated the conductivity of the discharge product in Li-O$_2$ cells to be $10^{-12}$-$10^{-13}$ S/cm based on electrochemical discharge/charge curves.[8] However, caution should be exercised in directly comparing these measurements to our calculations on crystalline Li$_2$O$_2$. First, the experiments were carried out at low capacities nominally resulting in Li$_2$O$_2$ deposits thin enough ($< 5$ nm[8,5]) to support electron tunneling.[5] Second, it is well known that side reactions[3,1,4,52] can alter the composition (and presumably the conductivity) of the experimental discharge product, and also contribute to the observed current density.[53,54] Finally, morphological features in the experimental deposits (surfaces, grain boundaries, interfaces, amorphous regions, $etc.$)[17,16,55] may participate in transport, and these effects are not included in the present study.

**Influence of exchange-correlation functional**

A recent DFT+U study ($U = 6$ eV) also reported hopping barriers comparable to the present values (0.39 to 0.48 eV), and noted that the barrier values were sensitive to the choice of $U$.[19] As the mixing parameter $\alpha$ in hybrid functionals is somewhat analogous to the $U$ parameter in DFT+U, we likewise expect that the hopping barrier will also depend upon the choice of $\alpha$. This is demonstrated in Figure 3, which compares the energy profiles obtained with the two values of the mixing parameter explored: $\alpha = 0.25$ ($i.e.$ the HSE06 functional[20,21]) and 0.48. The HSE06 calculation yields much smaller barriers of 38 and 143 meV, in good agreement with Ong $et\ al.$, who found barriers of 68 and 152 meV using the same functional.[14] To test geometry effects, we also calculated the $\alpha = 0.48$ barrier using the $\alpha = 0.25$ geometry. This lowered the in-plane and out-of-plane barriers by only 78 and 88 meV, indicating that the difference in barrier height between functionals is largely due to electronic structure effects.

As previously described, our predictions for the concentrations and hopping barriers for charge carriers in Li$_2$O$_2$ are based on an optimized choice for the fraction of exact change, $\alpha$. Since other choices for $\alpha$ are possible, it is important to examine the influence of the mixing parameter upon polaron energy levels and their (hopping) transition states. Figure 4 shows the energy levels (dashed lines) of the hole and electron polaron states, as determined from their formation energies referenced to the average electrostatic potential.[39,56] Three different values of $\alpha$, corresponding to increasing amounts of exact exchange, are considered: 0, 0.25, and 0.48. [The $\alpha = 0$ case corresponds to the semilocal PBE GGA functional (*i.e.* no exact exchange), $\alpha = 0.25$ corresponds to the HSE06 functional, and $\alpha = 0.48$ corresponds to the functional that reproduces the average $Li_2O_2$ bandgap predicted $G_0W_0$ and self-consistent GW calculations (see prior discussion).] In systems where the atomic geometry and wavefunction do not change with $\alpha$, the functional form of the HSE family[20,56] dictates that the energy will vary linearly with $\alpha$. [Deviations from linearity indicate the degree to which the wavefunction (and geometry, if the atom coordinates are relaxed) is changing.] If the wavefunction and geometry are fixed, increasing amounts of exact exchange will increasingly penalize partially occupied orbitals;[57] that is, configurations with partially occupied orbitals should become higher in energy with increasing $\alpha$. If the "correct" value of $\alpha$ is chosen, the penalty on partially occupied orbitals will exactly compensate for the self-interaction error from the semilocal exchange contribution.

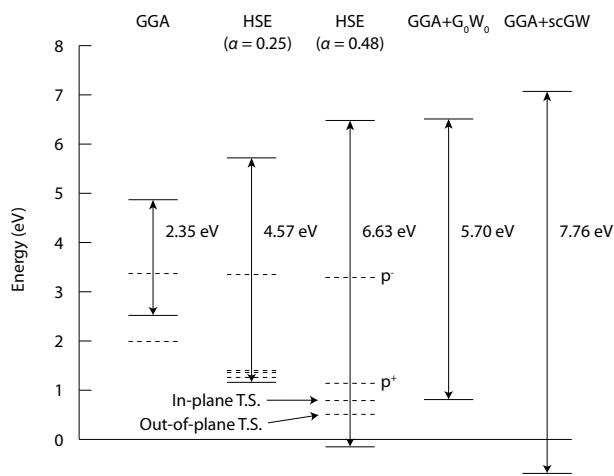

**Figure 4.** Energy levels associated with the band edges, polaron ground states, and transition states for polaron hopping in $Li_2O_2$ as a function of calculation method. Energies are referenced to the average electrostatic potential, which is assigned a value of zero. All energies were calculated using the $\alpha = 0.25$ geometries, and finite size corrections were not included. Transition states are not shown for the PBE functional

The band edges for the three functionals, as well as those obtained with GW methods[22,23] are shown as solid lines in Figure 4. Given that $G_0W_0$ and scGW band gaps typically bound the experimental band gap,[58,23] we expect that the positions of the $G_0W_0$ and scGW band edges likewise bound the positions of the experimental band edges. (Extra effort was taken to ensure convergence of the GW band edge positions, as these typically converge more slowly than the band gap;[39] see Figure S2 in the Supplementary Information.) Figure 4 shows that the valence band edge falls while the conduction band edge rises as $\alpha$ increases. This is expected given that the valence and conduction states involve the partial depletion/filling of molecular orbitals.[57] Note that this variation is essentially linear in $\alpha$, indicating that the valence and conduction band wavefunctions do not depend upon the choice of mixing parameter. The fact that the HSE06 valence band edge lies outside the range bounded by the GW edges suggests that a mixing parameter of $\alpha = 0.25$ is not sufficient to compensate the self-interaction error in $Li_2O_2$. On the other hand, a mixing parameter of $\alpha = 0.48$ places the valence band edge in better agreement with the GW calculations, indicating that this value gives a more realistic description of the electronic structure of $Li_2O_2$.

The data presented in Figure 4 illustrates a fundamental difference regarding the stability of hole polarons in $Li_2O_2$ as described by the semi-local PBE ($\alpha = 0$) *vs.* hybrid functionals ($\alpha = 0.25, 0.48$). In both hybrid functionals the position of the valence band maximum (VBM) lies below the hole polaron level. In contrast, the hole polaron level lies above the VBM in PBE. Consequently, charge depletion in PBE generates delocalized holes in the top of the valence band, whereas localized holes (polarons) are predicted by the hybrid functionals. (In order to make an apples-to-apples comparison, the energy levels in Figure 4 were determined using single-point energy calculations performed on the $\alpha = 0.25$ geometries. Releasing this constraint in PBE results in delocalization of the hole throughout the cell.) By comparing the PBE band edges to the GW band edges we can see that this instability is an artifact of self-interaction error.[38,39,57] This behavior is consistent with that of defects in other systems where semilocal functionals predict delocalized electrons, in contradiction to experimental measurements.[37,38]

Although PBE favors delocalized holes over hole polarons, Figure 4 shows that the hole polaron is actually more stable in PBE than in the hybrid functionals when referenced to the average electrostatic potential. This is because as the mixing parameter is reduced the hole polaron begins to spread out and hybridize with the valence band, resulting in partial occupancies of the oxygen p states and consequently a ground state energy that is too negative; see Figure S1 in the Supplementary Information. Although the energy levels in Figure 4 show that HSE06 ($\alpha = 0.25$) favors hole polarons over delocalized holes, the difference in energy between these two may be smaller than errors associated with finite-size effects and numerical convergence (see Figure S6 in the Supplementary Information); this raises some doubt as to the relative stability of delocalized holes and hole polarons in HSE06.[14]

As an aside, we note that the self-interaction errors inherent to GGAs are not limited to charged defects. Consider the neutral lithium vacancy, $V_{Li}^0$. The hybrid functionals predict this to consist of a $V_{Li}^--p^+$ bound pair, whereas PBE instead delocalizes the hole over several nearby oxygen sites. The resulting partial occupancy of oxygen p states and concomitant self-interaction error causes PBE to overbind this defect by as much as 1 eV relative to the hybrid functionals (see Supplementary Information). Indeed, a prior study using a GGA functional found a formation energy for $V_{Li}^0$ of 2.85 eV (referenced to bulk metallic Li), while a subsequent study using HSE06 found higher formation energies of 3.8 and 4.1 eV (TP and O sites, respectively). Our $\alpha = 0.48$ calculations yield similar values when referenced to metallic Li (3.98 and 4.33 eV).

Regarding the energy barriers for polaron hopping, we note that these transition states exhibit partial occupancy because the polaron is split between two different sites. Consequently, the energy levels of the transition states are sensitive to the choice of mixing parameter. Figure 4 illustrates the energy levels of the transition states for the in-plane and out-of-plane hopping pathways given in Fig. 3. This analysis also explains the variation of the hopping barrier with the choice of *U*, which also penalizes partially occupied orbitals.[19] As discussed above, the HSE06 mixing parameter of 0.25 is not large enough to compensate for self-interaction errors in $Li_2O_2$. The agreement with DFT+U hopping barrier[19] (over the optimal range of U values based on experimental data) lends additional support to our choice of mixing parameter, $\alpha = 0.48$. Furthermore, our preliminary calculations based on Marcus theory[59] yield a similar value for the in-plane barrier.

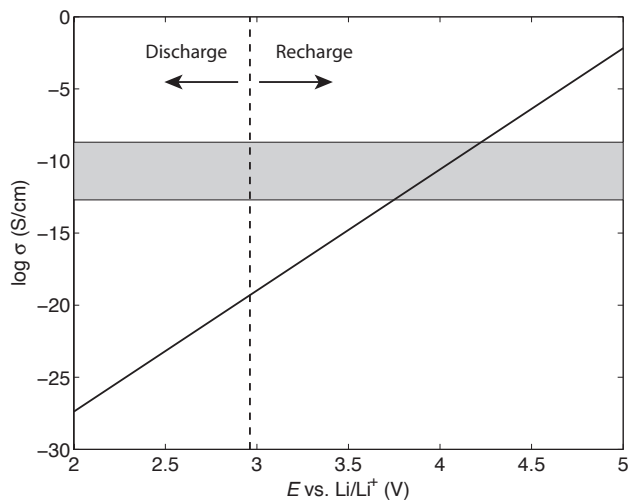

**Figure 5.** Predicted electronic conductivity as a function of cell voltage. The dashed line indicates the open circuit voltage. The gray shaded region indicates the target conductivity needed to meet performance requirements, as discussed in the text.

## Discussion

It is important to consider how the predicted conductivity could impact the performance of a Li-O$_2$ cell. To this end, we compare against performance targets for Li-air cells suggested in the literature.[60] We assume parameters based on the hypothetical bipolar plate-type Li-air battery described by Karulkar and Adams,[60] with the additional assumption that the discharge product grows as a uniform film on a porous cathode with a specific surface area of 100 m$^2$/g. Based on these assumptions, the discharge product should have a conductivity of $\sim 2 \times 10^{-11}$ S/cm in order to achieve an *iR* drop of less than 0.1 V (see Supplementary Information). This target value is several orders of magnitude larger than the predicted intrinsic electronic conductivity ($5 \times 10^{-20}$ S/cm), suggesting that charge transport through bulk (crystalline) $Li_2O_2$ can be a performance-limiting factor. We note that the migration of negative lithium vacancies cannot sustain charge transport over long time periods because the cathode materials used in Li-O$_2$ cells (typically porous carbon) are effectively ion blocking.[61] For this reason we focus on the electronic conductivity provided by hole polaron hopping.

### Discharge

As the predicted conductivity of $Li_2O_2$ is much smaller than that of other battery materials[47] it is tempting to conclude that charge transport through bulk $Li_2O_2$ is too small to play a meaningful role in a real cell. However, the conductivity is in principle not a fixed quantity, but can vary during discharge and charge because the cell potential impacts defect concentrations through variations in the lithium chemical potential. Figure 5 shows the predicted electronic conductivity as a function of cell voltage *E*. The conductivity increases exponentially with *E* because higher potentials favor delithiation (*i.e.* the creation of negative lithium vacancies, which are charge compensated by hole polarons). Under discharge conditions ($E < E_{OCV}$) the bulk electronic conductivity is far below the target value, and therefore unable to supply significant charge transport. The fact that fairly high capacities and discharge product sizes are obtained in experiments[10,11] suggests two possibilities: (*i.*) morphological features may locally enhance the conductivity of the discharge product; (*ii.*) the oxygen reduction reaction (ORR) is not occurring at the $Li_2O_2$ surface, but rather at the carbon support or catalyst.

Figure 6 summarizes the possible discharge mechanisms graphically. Figures 6a-c show mechanisms in which the ORR occurs at the $Li_2O_2$ surface and would therefore require charge transport through the discharge product. Electron tunneling (Figure 6a) cannot provide appreciable currents beyond a deposit thickness of ~5 nm,[5] and so this mechanism can occur only during the growth of thin deposits. Likewise, intrinsic conductivity (Figure 6b) is predicted to be quite low under discharge conditions, as discussed above. Therefore neither tunneling nor bulk conduction can account for the observed growth of large discharge product particles.[10,11] If charge transport does occur through $Li_2O_2$ during discharge, we expect that it must be along extended defects such as surfaces,[17] interfaces,[16] grain boundaries, dislocations, or amorphous regions[55] that can enhance conductivity (Figure 6c).

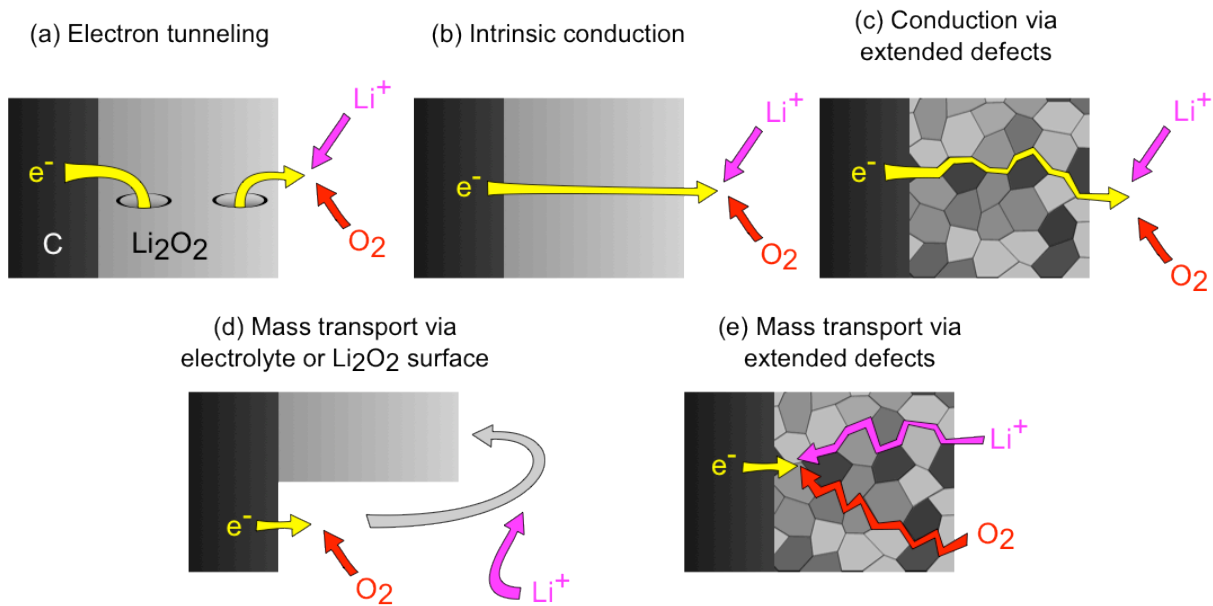

**Figure 6.** Possible discharge mechanisms for a Li-air cell.

Figures 6d-e illustrate mechanisms in which the ORR occurs not at the $Li_2O_2$ surface but rather at the surface of the carbon electrode (or catalyst). Figure 6d shows a scenario in which reduction happens at an exposed region of the carbon/catalyst, followed by diffusion through the electrolyte or along surfaces of existing $Li_2O_2$. In this case the deposition of the discharge product is not an electrochemical step, but a chemical step.[10,55] Figure 6e illustrates a more exotic scenario in which the ORR occurs at the buried carbon/$Li_2O_2$ interface; in this case, the reactants presumably diffuse through $Li_2O_2$ grain boundaries or other extended defects. Further investigation will be needed to explore these possibilities.

**Recharge**

Turning our attention to recharge, Figure 5 demonstrates that recharge conditions are more conducive to charge transport compared to discharge. That is, for each 119 mV of charging overpotential the conductivity increases by one order of magnitude, such that a 0.5 V recharge overpotential would enhance the conductivity by $2 \times 10^4$, and a 1 V overpotential would enhance it by a factor of $3 \times 10^8$, bringing the intrinsic electronic conductivity close to the targeted values (grey region in Figure 5). This effect results from an increase in the concentration of $p^+$ and $V_{Li}^-$ charge carriers at higher potentials. These results suggest that hole polaron hopping may be rapid enough to account for the observed rechargeability of bulk $Li_2O_2$ particles at moderately high potentials.

Our prediction that fairly large overpotentials are needed to activate charge transport is in qualitative agreement with the high (3.5 to 4.2 V), yet relatively flat potential profiles obtained upon the charging of cathodes packed with purchased $Li_2O_2$ powders.[24–27] On the other hand, much lower potentials have been observed upon the initial charging of cells with thin films of $Li_2O_2$;[9] in this case charge transport can proceed via electron tunneling.[9,5] Thus these two morphologies apparently have very different recharge profiles. (We note that impurities in the reference $Li_2O_2$ samples could also influence charging behavior.[25]) Recent experiments have demonstrated that Li-$O_2$ cells can concurrently form both thin *and* thick deposits.[62,55] By combining the electron tunneling narrative with our prediction of enhanced polaronic conductivity at higher potentials we arrive at the following two-stage process linking charge transport, particle morphology, and overpotentials during recharge, Figure 7. Charging will initiate at low potentials due to the dissolution of thin $Li_2O_2$ deposits or decomposition at/near the $Li_2O_2$/electrolyte/carbon three-phase boundary. Charging will then conclude at high potentials where thick deposits decompose via polaron hopping. Side reactions involving the electrolyte or carbon support may of course introduce further complications.[3,1,4]

In support of the above mechanism, we note that experiments involving electrolyte/cathode combinations that minimize side reactions (such as DMSO+LiClO$_4$/nanoporous gold[5] or CH$_3$CN+LiBF$_4$/P50[54]) yield charging profiles that start at low overpotentials and then rise to a plateau at overpotentials of roughly 1 V. Additionally, a recent study demonstrated a potassium air cell that discharged to potassium superoxide (KO$_2$), which is known to have a quite high room-temperature conductivity.[63] The low charging overpotentials observed in this experiment are consistent with the notion that sluggish charge transport in $Li_2O_2$ contributes to the recharge overpotentials in Li-$O_2$ cells.

If correct, our proposed mechanism implies that charge transport limitations will require moderately high overpotentials to recharge thick $Li_2O_2$ particles, *even if side reactions can be avoided*. Since high capacities likely require the formation of thick deposits, a tradeoff appears to exist between achieving both high capacities and efficient charging. Consequently, the ability to maximize capacity while simultaneously minimizing charging overpotentials will likely require accessing alternative reaction mechanisms that bypass bulk charge transport. Examples of such mechanisms would appear as the reverse of the processes shown in Figures 6c-e.

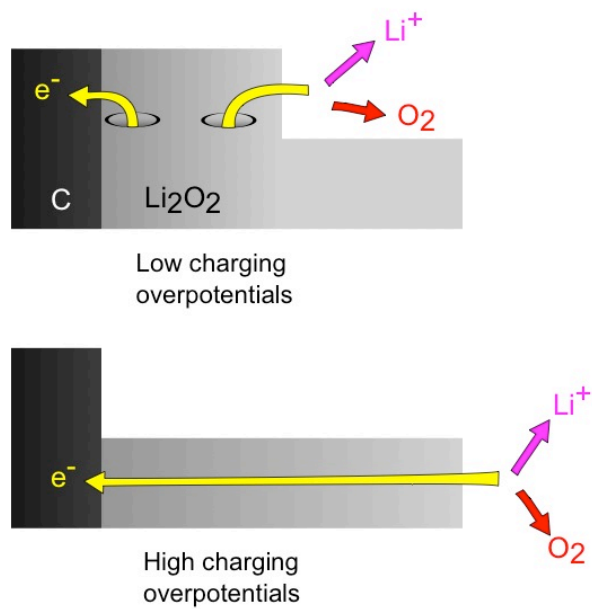

**Figure 7.** Proposed two-stage recharge mechanism for a Li-air cell.

We conclude by describing how the present results relate to other non-aqueous metal-air battery chemistries. More specifically, we speculate that the capability for even a modest amount of electronic charge transport in the discharge phase could explain why some non-aqueous metal-air chemistries are rechargeable at moderate potentials, while others are not. For example, $Li_2O_2$,[64–67] $Na_2O_2$,[68] $NaO_2$,[69] and $KO_2$[70] can be electrochemically decomposed in non-aqueous environments; on the other hand $Li_2O$ and $SiO_2$ are apparently electrochemically inactive in this context.[67,71–73] To rationalize these differences we recall that ionic solids in which the valence state can change tend to exhibit electronic conductivity due to the presence of charge carriers.[43,61] Examples include transition metal oxides in which the cation species can change its valence state (*e.g.* $TiO_2$ or $ZnO$[61,43]). This differs, of course, from the behavior in peroxides and superoxides where it is the anion that can change valence state. The results presented here suggest that the presence of $O_2$ dimers in $Li_2O_2$, $Na_2O_2$, $NaO_2$, and $KO_2$ may contribute to the rechargeability of these materials in non-aqueous metal-air batteries by providing a pathway for charge transport through the bulk. Although in $Li_2O_2$ overpotentials are needed to activate charge transport, in other compounds such overpotentials may not be required to achieve sufficient conductivity. For example, (as previously mentioned) potassium superoxide is known to exhibit a high room temperature conductivity.[63]

Lending further support to this hypothesis, recent Raman and magnetic measurements have provided evidence of superoxide ions in the $Li_2O_2$ discharge product[74] as well as synthesized $Li_2O_2$.[50] The presence of these ions confirms that $O_2$ dimers in $Li_2O_2$ can change their charge state between -2 and -1. In addition, prior calculations on surfaces[17,18] and on $Li_2O_2$ clusters[75] have identified superoxide-like dimers, and some alkali and alkaline earth metals are known to form mixed valence compounds in which peroxide and superoxide ions coexist.[76] In contrast, the absence of a species that can change valence state in $Li_2O$ and $SiO_2$ may account for the electrochemical inertness of these materials. For example, prior simulations and experiments have found that intrinsic conduction in $Li_2O$ is mediated by cationic Frenkel defects (*i.e.* $Li_i^+$ and $V_{Li}^-$),[77,78] and we do not expect the ionic conductivity associated with these defects to contribute to significant charge transport during cell operation because, as discussed above, the electrodes are ion-blocking.

## Conclusions

In summary, we have performed a detailed analysis of charge transport mechanisms in the primary discharge product of non-aqueous Li-air batteries, $Li_2O_2$. We observe that the defect chemistry of $Li_2O_2$ is driven by the ability of $O_2$ dimers to change valence state, and our calculations predict that a cation deficiency in $Li_2O_2$ is charge-compensated by small hole polarons. The intrinsic electronic and ionic conductivities of $Li_2O_2$ are predicted to be comparable (~$10^{-19}$ S/cm); the low bulk conductivity is predicted to limit the performance of Li-air cells. An enhancement of conductivity at high potentials where delithiation occurs may explain why bulk $Li_2O_2$ can nevertheless be decomposed electrochemically. Regarding computational methods, we find that the inclusion of exact exchange is essential for achieving a correct description of hole polarons, and that care must be exercised in the choice of mixing parameter as this can have a large impact on the polaron hopping barrier.

We propose that recharge in Li-$O_2$ cells may occur by a two-stage process, with thin deposits decomposing at low potentials and thick deposits decomposing at high potentials. Such a mechanism implies that it will be challenging to achieve both high capacities and high round-trip efficiency simultaneously: attaining high capacities requires the formation of large/thick $Li_2O_2$ deposits, and we argue that decomposing these deposits requires substantial overpotentials during charging. Consequently, techniques for accessing alternative mechanisms that bypass bulk charge transport should be explored.

The presence of species that can change its charge state may provide an important pathway for charge transport, and we propose that this feature explains why compounds containing $O_2$ dimers can be electrochemically decomposed in non-aqueous metal-air cells. This has implications for the development of other non-aqueous metal-air chemistries: for cations that cannot change charge state (e.g. Li, Na, K, Mg), only peroxide and superoxide discharge products (and not oxides) would be expected to be rechargeable. On the other hand, transition metals that can change valence state in principle may yield rechargeable non-aqueous metal-air chemistries even if the discharge product is an oxide.

## Acknowledgements

This work was supported by Robert Bosch LLC through the Bosch Energy Research Network Grant No.19.04.US11 and the U.S. Department Energy's U.S.-China Clean Energy Research Center for Clean Vehicles, grant no. DE-PI0000012. P. Albertus, B. Kozinsky, G. Samsonidze, and C. Monroe provided helpful comments on a preliminary version of this manuscript.

# Supplementary information for "Charge transport in lithium peroxide: relevance for rechargeable metal-air batteries"

Maxwell D. Radin[†], Donald J. Siegel*[‡]

[†]Department of Physics, University of Michigan, Ann Arbor, Michigan 48109, United States

[‡]Department of Mechanical Engineering, University of Michigan, Ann Arbor, Michigan 48109, United States

*E-mail: djsiege@umich.edu

**Defects**

Table S1 shows the calculated equilibrium formation energies for all defects studied using both PBE GGA and HSE hybrid functionals. Geometries were fully relaxed with each functional. Charge states of -1, 0, and +1 charge states were considered for all defect types. In cases where chemical intuition suggested that other charge states might be reasonable, we included additional possibilities. Given that oxygen atoms occur as covalently bonded dimers, we also considered oxygen divacancies ($V_{O2}$) in which an entire $O_2$ unit is removed. The initial guess used for the hole polaron geometry was created by shortening the O-O bond length by 10% and moving the nine nearest Li sites away from the O-O midpoint; six of the nine were moved 5% further away and three were moved 4%. The initial guess for electron polarons simply consisted of a 20% increase in the O-O bond length. While not included in the table below, we did consider hole bipolarons, $p^{2+}$. However, we were not able to localize both holes at a single $O_2$ dimer. Seven interstitial sites were considered: centers of empty prisms, centers of tetrahedra, centers of faces shared by prisms and octahedra, sides of prisms, centers of faces shared by tetrahedra and prisms, centers of faces shared by octahedra and tetrahedra, and off-center of empty prisms (displaced halfway towards a corner). GGA calculations were performed at all seven sites for all charge states considered; hybrid calculations were done using only the site predicted to be lowest in energy by GGA for a given charge state. For all charge states, oxygen interstitials were found to favor the site off-center of empty prisms, forming a structure that resembles an ozonide ion. Neutral lithium interstitials also favored this site, while the negative and positive lithium interstitials favored the faces of octahedra/tetrahedra and centers of empty prisms, respectively.

Figure S1 shows the magnetization density of the hole polaron. The distribution of the magnetization density represents the location of the hole. As $\alpha$ goes from 0 to 0.25 to 0.48, the hole becomes more localized.

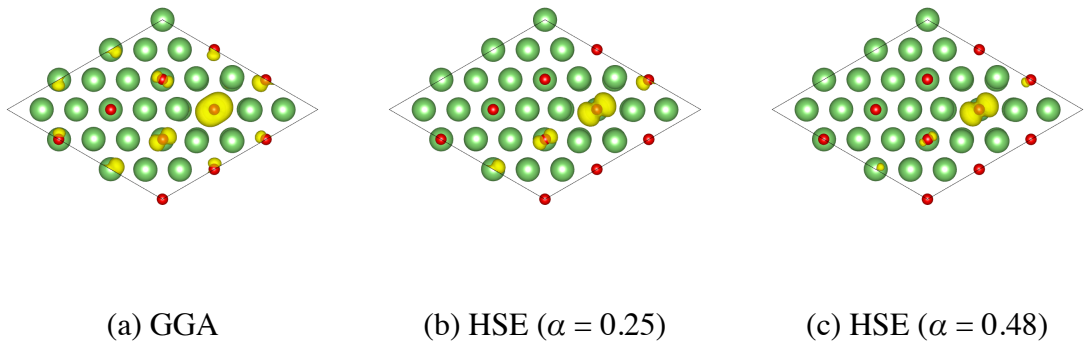

(a) GGA  (b) HSE ($\alpha = 0.25$)  (c) HSE ($\alpha = 0.48$)

**Figure S1.** Magnetization density isosurface of the hole polaron calculated with (a) PBE, (b) HSE ($\alpha = 0.25$), and (c) HSE ($\alpha = 0.48$). All three calculations shown here were performed at the $\alpha = 0.25$ geometry. The polaron is viewed from along the c axis.

**Table S1.** Equilibrium defect formation energies (eV) and concentrations (cm$^{-3}$) for all defects examined.

| | GGA | HSE ($\alpha = 0.25$) | HSE ($\alpha = 0.48$) |
|---|---|---|---|
| **p$^{2-}$** | 2.38 (3 × 10$^{-18}$) | 3.20 (6 × 10$^{-32}$) | 3.12 (1 × 10$^{-30}$) |
| **p$^{-}$** | 0.91 (1 × 10$^{7}$) | 1.44 (2 × 10$^{-2}$) | 1.51 (1 × 10$^{-3}$) |
| **p$^{+}$** | – | 0.85 (5 × 10$^{8}$) | 0.95 (1 × 10$^{7}$) |
| **V$_{Li}^{-}$ (O)** | 0.36 (3 × 10$^{16}$) | 0.87 (8 × 10$^{7}$) | 0.95 (3 × 10$^{6}$) |
| **V$_{Li}^{-}$ (TP)** | 0.29 (4 × 10$^{17}$) | 0.83 (4 × 10$^{8}$) | 0.93 (7 × 10$^{6}$) |
| **V$_{Li}^{0}$ (O)** | 0.37 (2 × 10$^{16}$) | 1.20 (2 × 10$^{2}$) | 1.37 (4 × 10$^{-1}$) |
| **V$_{Li}^{0}$ (TP)** | 0.22 (6 × 10$^{18}$) | 0.86 (1 × 10$^{8}$) | 1.02 (2 × 10$^{5}$) |
| **V$_{Li}^{+}$ (O)** | 0.81 (9 × 10$^{8}$) | 1.82 (9 × 10$^{-9}$) | 2.05 (1 × 10$^{-12}$) |
| **V$_{Li}^{+}$ (TP)** | 0.59 (4 × 10$^{12}$) | 1.23 (7 × 10$^{1}$) | 1.45 (1 × 10$^{-2}$) |
| **V$_{O}^{-}$** | 2.54 (1 × 10$^{-20}$) | 3.27 (8 × 10$^{-33}$) | 3.58 (4 × 10$^{-38}$) |
| **V$_{O}^{0}$** | 0.73 (3 × 10$^{10}$) | 0.74 (2 × 10$^{10}$) | 0.74 (2 × 10$^{10}$) |
| **V$_{O}^{+}$** | 1.16 (2 × 10$^{3}$) | 1.57 (3 × 10$^{-4}$) | 1.66 (9 × 10$^{-6}$) |
| **V$_{O2}^{-}$** | 1.81 (1 × 10$^{-8}$) | 2.38 (4 × 10$^{-18}$) | 2.47 (1 × 10$^{-19}$) |
| **V$_{O2}^{0}$** | 4.42 (2 × 10$^{-52}$) | 4.62 (7 × 10$^{-56}$) | 4.71 (2 × 10$^{-57}$) |
| **V$_{O2}^{+}$** | 4.70 (3 × 10$^{-57}$) | 4.34 (5 × 10$^{-51}$) | 4.32 (1 × 10$^{-50}$) |
| **V$_{O2}^{2+}$** | 4.24 (2 × 10$^{-49}$) | 3.35 (2 × 10$^{-34}$) | 3.24 (9 × 10$^{-33}$) |
| **O$_{i}^{2-}$** | 3.38 (2 × 10$^{-34}$) | 4.44 (3 × 10$^{-52}$) | 4.55 (4 × 10$^{-54}$) |
| **O$_{i}^{-}$** | 2.79 (1 × 10$^{-24}$) | 4.07 (5 × 10$^{-46}$) | 4.34 (1 × 10$^{-50}$) |
| **O$_{i}^{0}$** | 1.37 (8 × 10$^{-1}$) | 1.37 (8 × 10$^{-1}$) | 1.33 (5 × 10$^{0}$) |
| **O$_{i}^{+}$** | 1.70 (3 × 10$^{-6}$) | 2.20 (1 × 10$^{-14}$) | 2.22 (5 × 10$^{-15}$) |
| **Li$_{i}^{-}$** | 3.39 (1 × 10$^{-34}$) | 3.83 (4 × 10$^{-42}$) | 3.80 (1 × 10$^{-41}$) |
| **Li$_{i}^{0}$** | 2.65 (3 × 10$^{-22}$) | 2.59 (3 × 10$^{-21}$) | 2.51 (6 × 10$^{-20}$) |
| **Li$_{i}^{+}$** | 2.40 (1 × 10$^{-18}$) | 1.83 (5 × 10$^{-9}$) | 1.69 (1 × 10$^{-6}$) |

## GW Convergence Tests

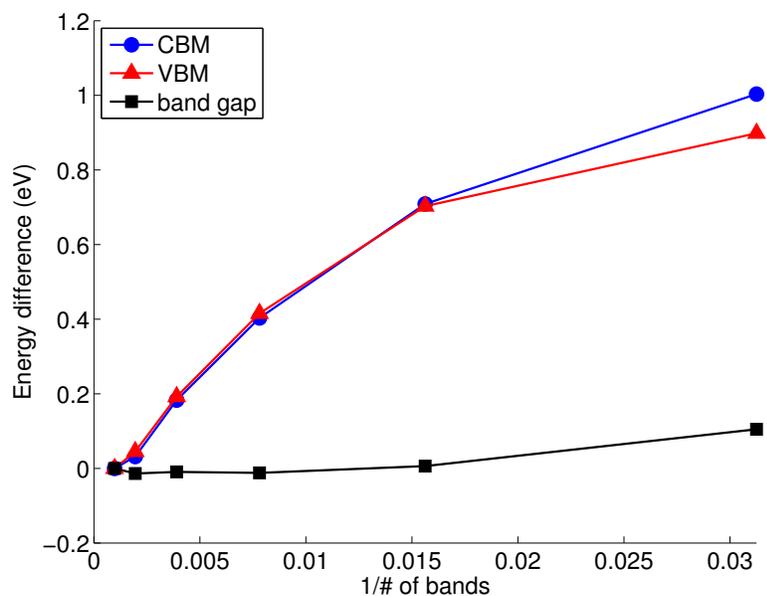

**Figure S2.** Convergence of the GGA+$G_0W_0$ band edges and band gap with respect to the number of bands used in the calculation. Data shown for calculations on a conventional unit cell with a number of bands equal to 32, 64, 128, 256, 512, and 1024. Based on this data, we chose to use 1024 bands. An extrapolation to an infinite number of bands indicates that the band edges are converged to within about 50 meV.

## Hopping and Migration Pathways

All five nearest neighbor $V_{Li}^-$ migration pathways were considered. The GGA nudged elastic band energy barriers for these paths are listed in Table S2 (see also Figure S3). For the A→D pathway, two values are given because the TP and O sites have slightly different energies.

**Table S1.** Migration barriers for $V_{Li}^-$ migration calculated using the nudged elastic band method at the GGA level of theory.

| Path | Barrier (eV) | Description |
| --- | --- | --- |
| A→B | 1.00 | In-plane between TP sites. |
| C→D | 1.06 | In-plane between O sites. |
| A→E | 2.34 | Out-of-plane between TP sites. |
| D→F | 1.60 | Out-of-plane between O sites. |
| A→D | 0.39/0.33 | Between TP and O sites. |

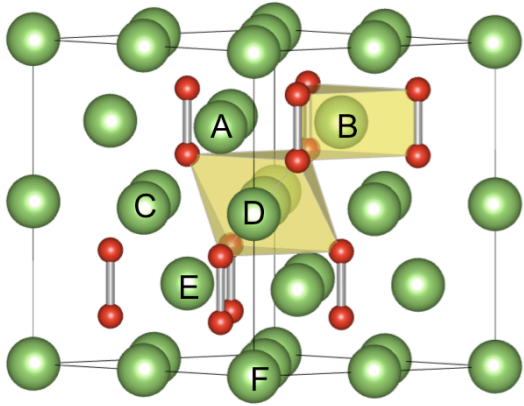

**Figure S3.** Lithium vacancy migration pathways.

Similarly, all nearest-neighbor hole polaron hopping pathways were considered. The $O_2$ dimers form a distorted hcp lattice in $Li_2O_2$. Neglecting the Jahn-Teller distortion of the hole polaron, there would be six symmetry equivalent in-plane nearest-neighbor hopping paths and six symmetry equivalent out-of-plane nearest-neighbor hopping paths. The Jahn-Teller distortion allows for three possible orientations of the initial and final states, so in principle there are 3 × 3

= 9 possible hopping paths between any two sites. However, several of these paths are symmetry equivalent. For in-plane hopping, there are only six symmetry inequivalent paths, and for out-of-plane hopping there are only four. Figure S4 summarizes these paths graphically and gives the calculated HSE barriers ($\alpha = 0.48$) based on a linear interpolation of images. Additionally, there is a trivial in-place rotation path, for which we find an barrier of 5 meV.

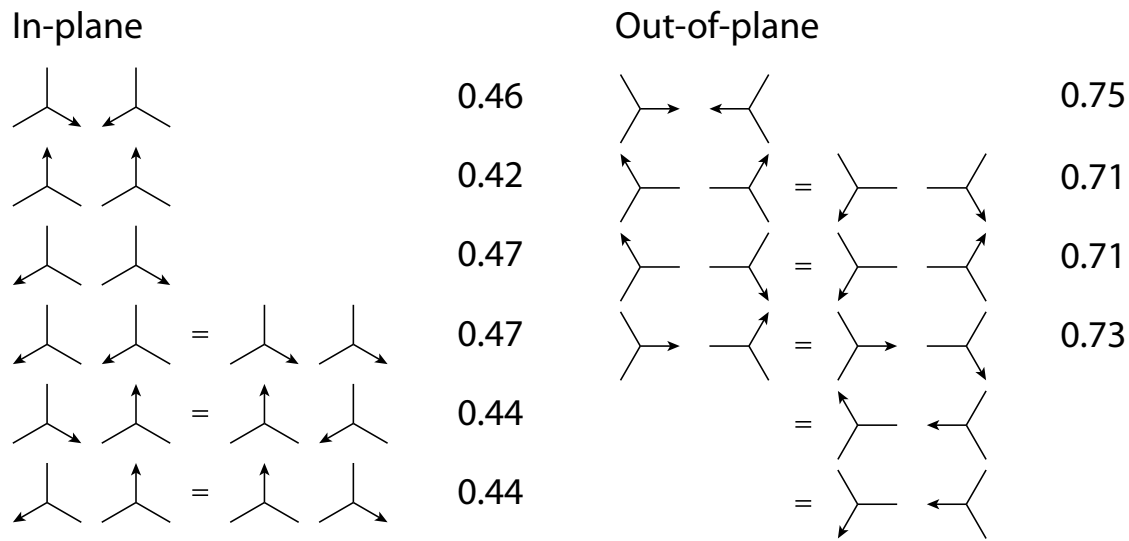

**Figure S4.** Nearest neighbor hole polaron hopping paths. Each path is depicted by two glyphs which represent the initial and final polaron states. The three lines represent the directions of three nearest trigonal prismatic Li sites. The arrow indicates the direction of the Jahn-Teller distortion (i.e. which of the three O-Li bonds is contracted). The hopping direction is left to right, and symmetry equivalent paths are indicated by an equals sign. Hopping barriers calculated from a linear interpolation of images are given in eV.

**Estimation of target conductivity**

We estimate the conductivity required for a hypothetical bipolar plate battery pack described by Adams and Karulkar.[1] We assume that the cathode uses carbon with a specific area of 100 m$^2$/g$_C$,[2] and that Li$_2$O$_2$ forms a film of uniform thickness. Based on the parameters shown in Table S2, the film will be of thickness $T = QV/4ae = 6$ nm, where $e$ is the elementary charge and the factor of four arises from the fact that four electrons are transferred per unit cell of $Li_2O_2$. The carbon loading is $L = E/NAUQ = 0.013$ $g_C/cm^2$, so the microscopic current density is $j = i/aL = 3.4$ µA/cm². To achieve an $iR$ drop across the discharge product of $\eta = 0.1$ V, the conductivity must be $\sigma = Tj/\eta = 2\times 10^{-11}$ S/cm. We assume an uncertainty of two orders of magnitude in this estimate.

**Table S2.** Parameters used to determine overpotential for a hypothetical Li-air battery.

| Parameter | Description | Units | Value |
|---|---|---|---|
| $Q$ | Specific capacity | C/$g_C$ | 1650 |
| $E$ | Pack energy | Wh | 40 |
| $N$ | Number of cells | Dimensionless | 1434 |
| $i$ | Macroscopic current density | mA/cm² | 42 |
| $A$ | Plate active area | cm² | 500 |
| $U$ | Cell voltage | V | 2.7 |
| $V$ | $Li_2O_2$ unit cell volume | Å³ | 64 |
| $a$ | Specific area of carbon | m²/$g_C$ | 100 |

**Computational Methods**

First principles calculations were performed using the Vienna *ab initio* simulation package (VASP),[3–6] with Blöchl's projector augmented wave (PAW) method,[7] 2s and 2s2p electrons treated as valence for Li and O, and a plane-wave cutoff energy of 400 eV. Occupancies were determined with a Gaussian smearing of width 0.1 eV for bulk phases and 0.01 eV for molecules. Lattice constants were determined by relaxing the cell shape, volume, and atomic positions of the conventional unit cells. All atomic forces were minimized to a tolerance of 0.02 eV/Å, with the cell shape/volume held constant during defect calculations. A 3×3×2 supercell (144 atoms) with Γ-point k-space integration was used for defect calculations. All calculations used the lattice constants obtained with $\alpha = 0.25$ ($a = 3.12$ Å and $c = 7.61$ Å). Since there has been no experimental measurement of the bandgap, we fit the mixing parameter $\alpha$ to the average of the GGA+$G_0W_0$ and GGA+scGW band gaps (calculated at the $\alpha = 0.25$ geometry); this choice is motivated by the fact that GGA+$G_0W_0$ is known to underestimate gaps, while GGA+scGW (in the absence of vertex corrections) overestimates gaps.[8,9] We found that a mixing parameter of $\alpha = 0.48$ reproduces the reference gap of 6.62 eV. Given the uncertainty in the true band gap, there is some uncertainty in the optimal value of $\alpha$ and therefore the polaron hopping barrier (and, to a lesser extent, defect formation energies). (Additionally the value of $\alpha$ that reproduces the true band gap may not exactly reproduce the true band edges nor the hopping barrier.[10]) A higher level of theory may be needed to further refine our estimate of the polaron hopping barrier.

The equilibrium concentration $C$ of a defect $X$ in charge state $q$ in a given solid phase can be written as $C(X^q) = D_X e^{-E_f(X^q)/k_B T}$, where $D_X$ is the number density of defect sites.[11] The formation energy $E_f$ is calculated as:[12]

$$E_f(X^q) = E_0(X^q) - E_0(\text{bulk}) - \sum_i n_i \mu_i + q\varepsilon_F + E_{\text{MP1}}$$

where $n_i$ is the number of atoms of the $i^{\text{th}}$ species in the defect, $\mu_i$ is the chemical potential of that species, $\varepsilon_F$ is the Fermi level, and $E_{\text{MP1}}$ is the Makov-Payne monopsize correction.[13,12]

We set the chemical potential of oxygen to be one half the free energy of gaseous $O_2$ at 300 K and 0.1 MPa; This condition captures a scenario under which the cathode and the electrolyte (including dissolved oxygen) are in equilibrium with oxygen in the air. We calculate the free energy of oxygen as

$$G(300\text{ K, }O_2) = E_0(O_2) + k_b T - TS_{\text{expt}}$$

where the $k_b T$ term accounts for the $pV$ contribution to free energy, and $S_{\text{expt}}$ is the experimental entropy.[14] We have intentionally neglected the small contributions to the free energy due to the translational, rotational, and vibrational degrees of freedom because we are not including these terms in the bulk phases; this maintains some degree of error cancellation. The chemical potential of Li in $Li_2O_2$ within the cathode may be related to the chemical potential of Li in the anode (BCC Li) via the following expression:[15]

$$\mu_{Li}(\text{cathode}) = \mu_{Li}(\text{BCC Li}) - eE$$

where $e$ is the elementary charge and $E$ is the cell potential. Note that the open circuit voltage corresponds to the same thermodynamic boundary condition as isolated $Li_2O_2$, $\mu_{Li}^{OCV} = \frac{1}{2}(G(300\text{ K}, Li_2O_2) - 2\mu_O)$. The assumption of local thermodynamic equilibrium within the cathode during cell operation ($E \neq E^{OCV}$) is justified by the facility of lithium kinetics over the length and time scales relevant to cell operation. The predicted $V_{Li}^-$ diffusivity[16] is fairly high at $6 \times 10^{-9}$ cm$^2$/s, which corresponds to a characteristic diffusion length[16] over one hour ($L = \sqrt{Dt} = 47$ μm) that is much larger than the typical discharge product particle size (1 μm or less[17,18]), and prior experimental and computation studies indicate that the kinetic barrier for lithium adsorption/desorption from a $Li_2O_2$ surfaces is quite low.[19–21]

Because DFT systematically overbinds gas-phase $O_2$ relative to solid oxides,[22,23,19] we correct the ground state energy of the $O_2$ molecule using the experimental formation enthalpy of $Li_2O_2$. For defect calculations, we apply a correction to the energy of $O_2$ based on the experimental formation enthalpy of $Li_2O_2$ at 300 K, $\Delta H_f(Li_2O_2)^{expt} = -6.57$ eV:[14]

$$E_0(O_2)^{corr} = E_0(O_2) + \Delta H_f(Li_2O_2)^{calc} - \Delta H_f(Li_2O_2)^{expt}$$

This increases the energy of $O_2$ molecule by 0.78, 0.68, and 0.58 eV for $\alpha = 0, 0.25$, and 0.48. We note that prior studies have found that the error in formation energy varies to some degree between different alkali and alkaline-earth metal oxides, peroxides, and superoxides.[19] This indicates that in addition to errors in the ground state energy of the $O_2$ molecule, there is some error associated with the solid phases. However, we note that our results are not greatly sensitive to the choice of correction: a 0.1 eV change in the $O_2$ correction changes the equilibrium hole polaron formation energy only by only 0.025 eV.

Lastly, we discuss finite-size effects in our simulations. While more complicated finite size corrections have been proposed, the monopole errors have been shown to be the leading error, scaling as one over the length of the supercell.[12] We note that the inclusion of the monopole correction is an improvement over previous studies on polarons in $Li_2O_2$, which did not include any finite size corrections.[24,25] Using density functional perturbation theory (with the PBE functional),[26] we have calculated the in-plane and out-of-plane relaxed-ion (*i.e.* low-frequency) dielectric constants of $Li_2O_2$ to be $\varepsilon_{xx} = \varepsilon_{yy} = 7.48$ and $\varepsilon_{zz} = 12.54$; given the relatively modest anisotropy, we simply adopt a value of $\varepsilon = 10$ for the purposes of calculating finite size corrections. This yields a correction of $E_{MP1} = 0.17$ eV for defects with a charge of $q = \pm 1$ in our 3×3×2 supercell.

Figure S5 shows that the MP1 correction significantly improves size convergence for the $V_{Li}^-$ (O) defect. We also performed some finite size tests on the hole polaron, as shown in Figure S6. However, because this defect is unstable in GGA, it was necessary to use a hybrid functional; consequently it was not possible to go to larger cell sizes. At small sizes, one can see that the hole polaron in HSE is more sensitive to supercell size than the negative lithium vacancy. Based on the magnetization density shown in Figure S1(b), we attribute this behavior to wavefunction overlap between periodic images.[27]

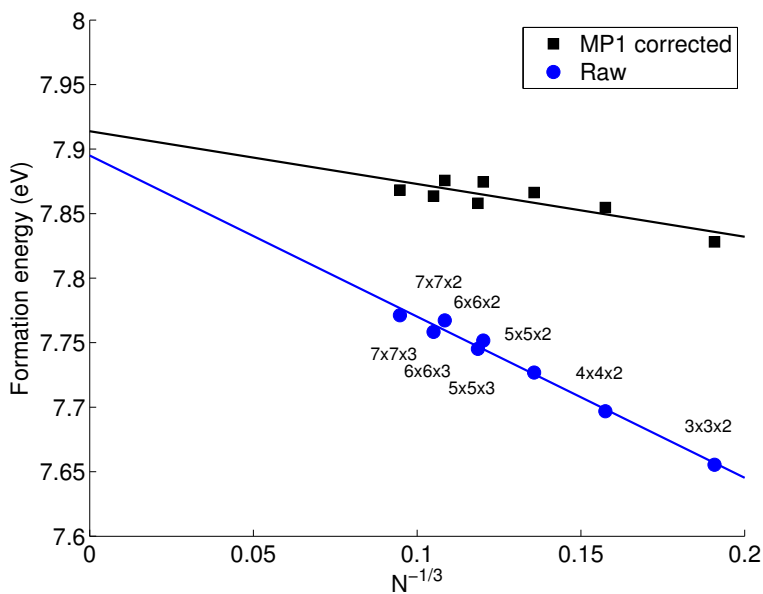

**Figure S5.** Size convergence of the $V_{Li}^-$ (O) GGA formation energy referenced to the average electrostatic potential. Calculations were performed up to a $7 \times 7 \times 3$ supercell (N = 1176 atoms). A linear fit is shown to allow for extrapolation to infinite supercell size.

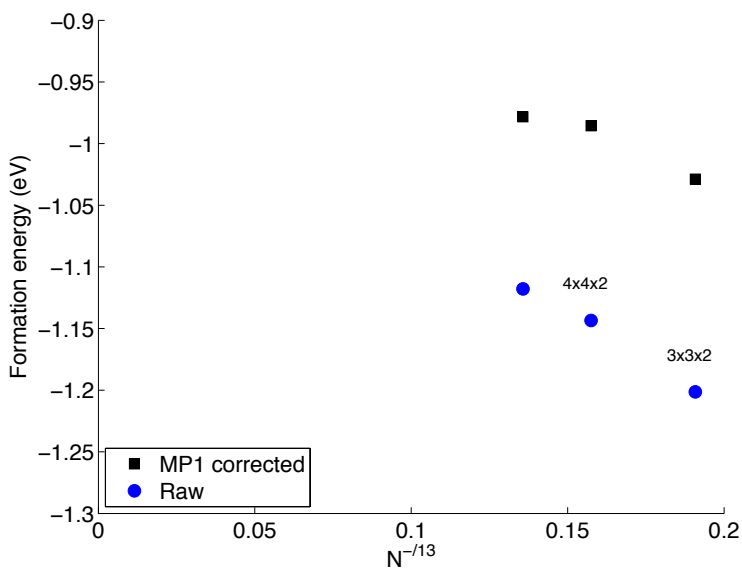

**Figure S6.** Size convergence of the hole polaron HSE ($\alpha = 0.48$) formation energy referenced to the average electrostatic potential. Calculations were performed up to a $5 \times 5 \times 2$ supercell (N =

400 atoms). We refrain from including a linear fit because the errors due to wavefunction overlap are not expected to have a linear dependence on the cell dimension.